\documentclass[review]{elsarticle}

\usepackage{lineno,hyperref}
\modulolinenumbers[5]

\usepackage{amstext,amsmath,amssymb}
\usepackage{times}
\usepackage{graphicx}
\usepackage{algorithmic}
\usepackage{algorithm}
\usepackage{url}
\usepackage{hyperref}
\usepackage{breakurl}
\usepackage{graphicx}
\usepackage{color}
\usepackage{verbatim}
\usepackage{subfig}
\usepackage{multirow}

\newcommand{\bdm}{
    \begin{displaymath}}

\newcommand{\edm}{
    \end{displaymath}}

\newcommand{\be}{
    \begin{equation}}

\newcommand{\ee}{
    \end{equation}}

\newcommand{\bea}{
    \begin{eqnarray}}

\newcommand{\eea}{
    \end{eqnarray}}

\newcommand{\beit}{\begin{itemize}}
\newcommand{\eeit}{\end{itemize}}

\newcommand{\eat}[1]{ }

\begin{document}

\begin{frontmatter}

\title{Exploring Mixed Integer Programming Reformulations for Virtual Machine Placement with Disk Anti-Colocation Constraints}

\author[firstaddress]{Xiaoying Zheng}
\ead{zhengxy@sari.ac.cn}

\author[secondaryaddress]{Ye Xia}
\ead{yx1@cise.ufl.edu}

\address[firstaddress]{Shanghai Advanced Research Institute, Chinese Academy of Sciences, China}
\address[secondaryaddress]{Department of Computer and Information Science and Engineering, University of Florida, Gainesville, FL 32611}

\begin{abstract}
One of the important problems for datacenter resource management is to place virtual machines (VMs) to physical machines (PMs) such that certain cost, profit or performance objective is optimized, subject to various constraints. In this paper, we consider an interesting and difficult VM placement problem with disk anti-colocation constraints: a VM's virtual disks should be spread out across the physical disks of its assigned PM. For solutions, we use the mixed integer programming (MIP) formulations and algorithms. However, a challenge is the potentially long computation time of the MIP algorithms. In this paper, we explore how reformulation of the problem can help to reduce the computation time. We develop two reformulations, by redefining the variables, for our VM placement problem and evaluate the computation time of all three formulations. We show that they have vastly different computation time. All three formulations can be useful, but for different problem instances. They all should be kept in the toolbox for tackling the problem. Out of the three, formulation COMB is especially flexible and versatile, and it can solve large problem instances.
\end{abstract}

\begin{keyword}
Datacenter, Cloud Computing, Virtual Machine Placement, Resource Management, Mixed Integer Programming
\end{keyword}

\end{frontmatter}



\section{Introduction}
\label{sec:introduction}

Cloud computing has gained firm traction in the marketplace as major high-tech companies rush to offer cloud services, such as Amazon AWS, Google AppEngine, Microsoft Azure, and Apple iCloud. For cloud providers, in order to get the best return for investment and to provide the best possible service to customers, one critical task is to manage the datacenter resources effectively. Today's resource management in datacenters involves a core problem, known as {\em virtual machine (VM) placement}. Each customer specifies a desired number of VMs, as well as the resource requirements for each VM, including CPU, memory, storage, I/O throughput, and possibly bandwidth between VM pairs \cite{XF10, WMZ11,JPX12}. A cloud provider's datacenters have a large number of physical machines (PM) mounted on racks and connected through layers of switches that form the datacenter network \cite{Cisco}. The VM placement problem is to assign the VMs to the PMs so that certain cost, profit or performance objective is optimized, subject to the PMs' resource capacity constraints and possibly network bandwidth constraints.

There is a great variety of VM placement problems, depending on what clouds offer, what customers need, and the performance/cost objectives of both parties. One category of services that customers often request contains {\em anti-colocation} requirements, which take the generic form that a set of requested resources should not be colocated in a sense that depends on the precise specification. For instance, to improve the availability of its service, a customer may require some of its VMs not to be placed on the same physical server or the same server rack \cite{AAS14}.

This paper focuses on a special type of anti-colocation requirements -- {\em disk anti-colocation}. Many VM types offered by public clouds such as Amazon EC2 \cite{EC2Inst} have multiple virtual disks per VM. When a customer requests such a VM, he may be interested in the following disk anti-colocation requirement: No physical disk of the PM to which the VM is assigned should contain more than one of the VM's virtual disks. That is, the VM's virtual disks should be spread out across the physical disks of the PM.



Our earlier paper \cite{XTFC17} has discussed the use cases and benefits of disk anti-colocation extensively. Here, we summarize that discussion. Cloud users often care a great deal about disk IO performance. Since local disks (or directly attached disks) to PMs have numerous advantages over network-based storage, such as higher IO throughput, lower latency, more predicable IO performance, lower cost and lower complexity \cite{AWS_Storage, Datadog}, they are the preferred storage option for many high-valued, critical applications such as NoSQL databases, Hadoop/MapReduce storage nodes, log or data-processing applications \cite{Datadog, ScyllaDB, DataStax, Hadoop_hw}. For such applications, when a requested VM is assigned to a PM, the VM's virtual disks will be mapped to the local physical disks of the PM. When disk anti-colocation is satisfied, accesses to different virtual disks do not interfere with each other; the users of the VM can expect improved disk IO performance, especially when RAID is used.

Although our problem adds only one complication -- disk anti-colocation -- to the classical VM placement problem, it is far more difficult to solve than the classical VM placement problem\footnote{In this paper, we do not focus on theoretical computation complexity, but on usable algorithms. The classical VM placement problem is a form of multi-dimensional bin packing problems. Even the basic one-dimensional bin packing problem is NP-hard \cite{CK04}.}. This greater difficulty can be seen later from the problem formulations, for instance, by counting the number of decision variables. It can also be seen intuitively. There are two levels of assignment to be made: One is to assign VMs to PMs; the other is to assign virtual disks to physical disks. What makes the overall problem especially difficult is that the two levels of assignment are intertwined. To the best of our knowledge, there are no known optimal combinatorial algorithms to solve the problem, other than naive enumeration (see Section \ref{sec:related} for detailed discussion).

We advocate the use of mixed integer programming (MIP) \cite{WN99} formulations and algorithms for our problem. The benefits of using MIP were argued in \cite{XTFC17}, and it has been used in a number of prior studies on similar resource management problems \cite{BCFF12, Fang2013, PLPMC13, RG17}. The MIP approach should complement other approaches that are frequently used for datacenter resource management, including specialized combinatorial algorithms and heuristic algorithms. 

The main challenge with the MIP approach is that the MIP algorithms can take a long time to find an optimal solution. Typical strategies to cope with that challenge include finding better algorithms, using more powerful computers to run the algorithms, or reformulating the problem differently. In this paper, we will explore the third strategy -- reformulation \cite{Trick05} -- to reduce the computation time for the disk anti-colocation problem. The paper presents three MIP formulations. The first one, F1, was the original formulation developed in \cite{XTFC17}; it is shown here for completeness and for comparison. The main contributions of this paper are in developing two additional formulations, F2 and COMB, and in evaluating and comparing the computation time of all three formulations. Formulations F2 and COMB involve non-obvious reformulation of the variables. That is, they define the variables very differently from what an obvious formulation does (in our case, formulation F1 is the obvious formulation). As Trick suggests, it is this type of reformulation that the modelers can make the most contribution in reducing the computation time because MIP solvers are not sophisticated enough to perform such reformulation \cite{Trick05}.

From our evaluation of the formulations, we arrive at the following main observations. Different formulations lead to drastically different computation time. However, which formulation has the least computation time depends on the problem instance. All three formulations can be useful for the right instances. But, formulation COMB is especially flexible and versatile, and it can solve large problem instances. Throughout the paper, we have discussions about how to decide which formulation to use in different situations.

Due to the inherent difficulty of our problem, when the problem size becomes large enough, no algorithm will be able to solve it optimally. In that case, one has to resort to non-optimal heuristic algorithm. Our earlier work \cite{XTFC17} explores how to solve large problem instances with a heuristic decomposition approach. The approach reported in this paper and the approach in \cite{XTFC17} are different but complementary.

The rest of the paper is organized as follows. In Section \ref{sec:related}, we discuss additional related work. In Section \ref{sec:model}, we describe our VM placement problem and present three MIP formulations. In Section \ref{sec:experiments}, we present experimental results to compare the computation time of the three formulations. Conclusions are given in Section \ref{sec:conclusion}. 

\section{Additional Related Work}
\label{sec:related}

There is a large body of research on different VM placement problems, such as VM placement with traffic awareness or network constraints \cite{MPZ10, AL12, BCFF12, ZYLW15}, with routing \cite{JLHCC12, PLPMC13}, with resource sharing by co-located VMs \cite{SSS11, RG15, RG17}, with energy awareness \cite{Fang2013, Li20131222, marotta2015simulated}, with random or time-varying resource requirements \cite{WMZ11, CS14, MSY12}.
Our earlier work \cite{XTFC17} and the follow-up paper by other authors \cite{Hbaieb2017} are the only papers that consider disk anti-colocation. In \cite{Hbaieb2017}, Hbaieb et al. propose a more scalable algorithm combining a decomposition method with local search heuristic. Neither paper deals with optimal algorithms.

Most VM placement problems, like ours, are superclasses of the vector bin packing problem, which is well-known to be NP-hard. Even for the vector bin packing problem, there have been relatively few exact (i.e., optimal) algorithms in the literature. Instead, research has focused on approximation algorithms and online algorithms (see \cite{CKPT17} for a recent survey). Within the exact algorithms, nearly all are about 1 or 2-dimensional vector packing with identical bins \cite{HDC94,DIM16}, whereas practical VM placement problems usually have more than two dimensions (i.e., resource types) and different bin (i.e., PM) types. More importantly, many VM placement problems like ours are more than vector bin packing. In our case, even if we have an exact algorithm for general vector bin packing with multiple bin types, it still won't solve our problem in which disk anti-colocation is coupled with vector bin packing.

A majority of prior studies on VM placement avoid MIP formulations all together. In the cases where MIP formulations are used, they are usually used to describe the problem; the algorithms are usually not based on MIP. Instead, the effort is usually on developing specialized combinatorial algorithms, such as multi-dimensional bin-packing heuristics or approximation algorithms \cite{AT07, CZS11, WMZ11}, graph algorithms \cite{MPZ10, AL12, ZYLW15} or other sophisticated heuristics \cite{Fang2013}. None of these are exact algorithms. For VM placement problems, changes to the problem specification often make the original algorithm inapplicable, unless it is a general MIP algorithm. The above algorithms are tailored to the special problems that the authors study, usually relying on certain structures of the problems. In our assessment, they cannot be adapted easily to our problem, due to the addition of the disk anti-colocation requirement, which poses difficult constraints of a different kind.

There is a small number of prior studies that do use MIP, but they consider very different problems from our problem \cite{Li20131222, BCFF12, Fang2013, PLPMC13,WNLL13, marotta2015simulated, RG17}. For instance, \cite{Fang2013} studies a problem of VM placement with energy-aware routing;  \cite{PLPMC13} studies a problem of placing customer-requested virtual networks into the datacenter's physical substrate, subject to the capacity constraints of physical nodes and physical links; \cite{RG17} formulates and solves an MIP problem for sharing-aware VM placement where colocated VMs can share memory pages. These earlier studies provide only one MIP formulation, but do not attempt problem reformulation.

Practical cloud systems usually adopt less sophisticated heuristics, such as round-robin, first-fit or first-fit-decrease, as evidenced by open-source middleware stacks \cite{cloudstack,openstack,euca}. While simple heuristics may find solutions quickly, they can also be underachieving in terms of performance. In particular, when a problem is sufficiently complex or have difficult constraints, intuitions that are needed to develop sound heuristics may fail. The anti-colocation constraints in our problem are difficult. It is not easy to design a heuristic algorithm that always has good performance.

\section{Three Problem Formulations}
\label{sec:model}

In this section, we present three MIP formulations of our VM placement problem and discuss their complexity and applicability. In the next section, we will evaluate their computation time when a standard MIP solver is used. In Table \ref{tab:notation}, we summarize the major notation.

\begin{table}[t]
\caption{Major Notation.}
\label{tab:notation}
\begin{center}
\begin{tabular}{c|c|c|c}
\hline
$N$  & number of VMs & $M$ & number of PMs  \\
$i$ & index of VM & $j$ & index of PM \\
$\mathcal{V}$ & the set of VMs & $\mathcal{P}$ & the set of PMs \\
$k$ & index of virtual disk & $l$ & index of physical disk \\
$\mathcal{T}^v$ & the set of VM types & $\mathcal{T}^p$ & the set of PM types \\
$u$ & index of VM type in $\mathcal{T}^v$ & $v$ & index of PM type in $\mathcal{T}^p$ \\
$t$ & a configuration ID  & $\mathcal{Y}_v$ & the set of all type-$v$ PMs \\
\hline
$C_j$ & \multicolumn{3}{|c}{the number of vCPUs that PM $j$ can support} \\
$M_j$ & \multicolumn{3}{|c}{the amount of memory (in GiB) of PM $j$} \\
$\hat{c}_j$ & \multicolumn{3}{|c}{a fixed cost associated with running PM $j$} \\
$\alpha_i$ & \multicolumn{3}{|c}{the number of vCPUs required by VM $i$} \\
$\beta_i$ & \multicolumn{3}{|c}{the memory requirement (in GiB) by VM $i$} \\
$R_i=\{1,\ldots, |R_i|\}$ & \multicolumn{3}{|c}{a set of virtual disks requested by VM $i$} \\
$\nu_{ik}$ & \multicolumn{3}{|c}{the requested disk volume size (in GB) for the requested virtual disk $k \in R_i$} \\
$D_j=\{1,\ldots, |D_j|\}$ & \multicolumn{3}{|c}{the set of available physical disks of PM $j$} \\
$S_{jl}$ & \multicolumn{3}{|c}{the size (in GB) of the physical disk $l \in D_j$} \\
$x_{ij}$ & \multicolumn{3}{|c}{the binary assignment variable from VM $i$ to PM $j$} \\
$y_{ikjl}$ & \multicolumn{3}{|c}{a binary assignment variable for disk} \\
$z_j$ & \multicolumn{3}{|c}{the binary variable indicating whether PM $j$ is used by some VMs} \\
$\mathcal{C}_v$ & \multicolumn{3}{|c}{the ID set of all the feasible configurations with respect to a type-$v$ PM} \\
$m_u$ & \multicolumn{3}{|c}{the total number of type-$u$ VMs that need to be placed} \\
$\textbf{w}^t$ & \multicolumn{3}{|c}{the vector representation of configuration $t$} \\
$w^t_u$ & \multicolumn{3}{|c}{the number of type-$u$ VMs in configuration $t$} \\
$\gamma_{jt}$ & \multicolumn{3}{|c}{a binary PM-to-configuration assignment variable} \\
$\mathcal{P}_2$ & \multicolumn{3}{|c}{the set of PMs with moderate numbers of feasible configurations}\\
$\mathcal{P}_1$ & \multicolumn{3}{|c}{$\mathcal{P}_1 = \mathcal{P} \backslash \mathcal{P}_2$} \\
$B$ & \multicolumn{3}{|c}{a sufficient large constant} \\
\hline
\end{tabular}
\end{center}
\end{table}





Consider $N$ VMs and $M$ PMs. Each VM has the following resource requirements: memory, number of vCPUs, number of local disk volumes (virtual ones) and their respective sizes. Each PM has certain memory capacity, number of vCPUs that it can support, and number of local disks and their respective sizes. These local disks may be in the PM or directly attached.

We first give an overview of the constraints for our problem.
\begin{itemize}
\item There are the usual capacity constraints for each resource: With respect to the vCPU or memory resource, the total amount of resource required by all the VMs assigned to a PM cannot exceed the resource capacity of the PM.

\item The next set of constraints is quite special, which makes our problem different from the usual VM placement problems. When multiple virtual disks are requested for a VM $i$, there is a disk {\em anti-colocation constraint}: No physical disk of the PM (to which VM $i$ is assigned) should contain more than one of VM $i$'s requested virtual disks. The motivations for such a constraint have been given in Section \ref{sec:introduction}.

\item A final set of constraints is that the aggregate size of all virtual disks assigned to a physical disk cannot exceed the capacity of the physical disk.

\end{itemize}

The optimization objective will ultimately be decided by the cloud provider. For concreteness, we assume that a fixed operation cost is incurred for a PM as long as the PM is used by some VMs (that is, some VMs are assigned to the PM). Specifically, when a PM $j$ is turned on to host some VMs, there is a fixed cost $\hat{c}_j$ associated with running the PM; when the PM is off, there is zero cost. The operation cost may include the average energy cost when a machine is running and typical maintenance cost. The optimization objective is to minimize the total operation cost of all the used PMs.


The model can be enriched in many ways. With respect to the costs and objective, we may include load-dependent costs in the optimization objective. For instance, the energy cost of a PM may be larger when the CPU load is higher. The model can also be extended to include local and network bandwidth constraints, although network constraints pose great difficulty and require additional techniques for solutions \cite{MPZ10, Fang2013}. Those additional constraints depend on actual customers' needs and cloud providers' policies, and they vary across customers/providers and change over time. Given the absence of details, we will not include those additional constraints in this paper. We expect that disk anti-colocation is a class of distinct constraints. It is worthwhile to single it out for a focused investigation.





\subsection{Formulation 1 - Direct Assignment}

Let the sets of the VMs and PMs be denoted by $\mathcal{V}$ and $\mathcal{P}$, respectively. For each VM $i$, let $\alpha_i$ be the number of vCPUs required and let $\beta_i$ be the memory requirement (in GiB).\footnote{1 GiB (gibibyte) is equal to $2^{30}$ bytes, which is $1,073,741,824$ bytes; 1 GB (gigabyte) is equal to $10^{9}$ bytes.} For each VM $i$, a set of virtual disks is requested and the set is denoted by $R_i=\{1,\ldots, |R_i|\}$. For each of the requested virtual disks $k \in R_i$, let $\nu_{ik}$ be the requested disk volume size (in GB).




For each PM $j$, let $C_j$ be the number of vCPUs it can support, $M_j$ be the amount of memory (in GiB), and $D_j=\{1,\ldots, |D_j|\}$ be the set of available physical disks. The sizes of the physical disks are denoted by $S_{jl}$ (GB) for $l \in D_j$.


For each $i \in \mathcal{V}$ and each $j \in \mathcal{P}$, let $x_{ij}$ be the binary assignment variable from VM $i$ to PM $j$, which takes the value $1$ if VM $i$ is assigned to PM $j$ and $0$ otherwise. The binary variables $y_{ikjl}$ are used for disk assignment: $y_{ikjl}$ is set to 1 if VM $i$ is assigned to PM $j$ and the requested virtual disk $k$, where $k \in R_i$, for VM $i$ is assigned to the physical disk $l$ of PM $j$, where $l \in D_j$; it is set to $0$ otherwise. Let $z_j$ be a 0-1 variable indicating whether PM $j$ is used by some VMs. The following is our first formulation for VM placement.






\begin{align}
\textbf{F1:}
\min_{x,y,z} & \sum_{j \in \mathcal{P}} \hat{c}_j z_j \label{eq:optobj_1} \\
\text{s.t.} \ \ & y_{ikjl} \leq x_{ij}, \ \ \ i \in \mathcal{V}, j \in \mathcal{P}, k \in R_i, l \in D_j \label{eq:diskconstonlyif} \\
& \sum_{j \in \mathcal{P}} \sum_{l \in D_j} y_{ikjl} = 1, \ \ \ i \in \mathcal{V}, k \in R_i \label{eq:mustassign} \\
& \sum_{j \in \mathcal{P}} x_{ij} = 1, \ \ \ i \in \mathcal{V} \label{eq:onePMonly} \\
& \sum_{k \in R_i} y_{ikjl} \leq 1, \ \ \ i \in \mathcal{V}, j \in \mathcal{P}, l \in D_j \label{eq:diskexclusive} \\
& \sum_{i \in \mathcal{V}} \sum_{k \in R_i} \nu_{ik} y_{ikjl} \leq S_{jl}, \ \ \ j \in \mathcal{P}, l \in D_j \label{eq:diskcap}
\end{align}
\begin{align}
& \sum_{i \in \mathcal{V}} \alpha_{i} x_{ij} \leq C_{j}, \ \ \ j \in \mathcal{P} \label{eq:corecap} \\
& \sum_{i \in \mathcal{V}} \beta_{i} x_{ij} \leq M_{j}, \ \ \ j \in \mathcal{P}. \label{eq:memcap} \\
& z_j \leq \sum_{i \in \mathcal{V}} x_{ij}, \ \ \ j \in \mathcal{P} \label{eq:z1}\\
& B z_j \geq \sum_{i \in \mathcal{V}} x_{ij}, \ \ \ j \in \mathcal{P} \label{eq:z2}\\
& x_{ij}, y_{ikjl}, z_j \in \{0,1\}, \ \ \ i \in \mathcal{V}, k \in R_i, j \in \mathcal{P}, l \in D_j. \nonumber
\end{align}
The following explains some of the constraints:
\begin{itemize}
\item (\ref{eq:diskconstonlyif}) ensures that the requested virtual disks for VM $i$ may be assigned to the physical disks of PM $j$ only if VM $i$ is assigned to PM $j$.

\item (\ref{eq:mustassign}) ensures that every requested virtual disk must be assigned to exactly one physical disk.

\item (\ref{eq:onePMonly}) ensures that every VM is assigned to exactly one PM.

\item (\ref{eq:diskexclusive}) ensures that VM $i$ cannot have more than one of its virtual disks assigned to the same physical disk; (\ref{eq:diskconstonlyif}) and (\ref{eq:diskexclusive}) together enforce the disk anti-colocation constraints.

\item (\ref{eq:diskcap}) is the disk capacity constraint.

\item (\ref{eq:corecap}) and (\ref{eq:memcap}) are the resource capacity constraints posed by the number of vCPUs and the total memory size of each PM $j$.

\item (\ref{eq:z1}) and (\ref{eq:z2}) together ensure that $z_j=1$ if and only if $x_{ij}=1$ for some $i \in \mathcal{V}$. In (\ref{eq:z2}), $B$ is a large enough constant (it is enough to take $B = N$).
\end{itemize}

\noindent  {\bf Remark.} The difficulty of our problem is reflected first by the $y_{ikjl}$ variables, which are indexed by four subscripts, implying a large number of such variables. Moreover, there are two levels of assignments, VM assignment and disk assignment, and (\ref{eq:diskconstonlyif}) implies that they cannot be separated.
Finally, in formulation F1, we assume that each active PM has a fixed cost. In reality, some cost may be load dependent. For instance, the energy cost of a PM may depend on the number of VMs assigned to it. If the load-dependent energy cost needs to be incorporated and if the energy cost depends on the load linearly, our model only requires a small modification: We only need to modify the objective function by adding a linear term in the $x$ variables. There will be no other changes to the constraints.


\subsection{Formulation 2 -- Assign Configurations}
\label{sec:F2form}

The numbers of VM and PM types are often much smaller than the total numbers of VMs and PMs, respectively. For example, in Amazon EC2 \cite{EC2Inst}, there are only 40 VM types. Amazon does not disclose the detailed configurations of their PMs. From \cite{EC2Inst}, one can deduce that the PMs falls into a small number of types. For example, the five m4-type VMs are all supported by 2.4 GHz Intel Xeon E5-2676 v3 (Haswell) processors; the five c4-type of VMs are all supported by Intel Xeon E5-2666 v3 (Haswell) processors\footnote{We will see in Section \ref{sec:analysis} that, for formulation F2 and formulation COMB to work, having a small number of VM types is more important than having a small number of PM types. With a small number of VM types, the dimension of the configuration vectors is small. The formulations can still be effective even if the number of PM types is in thousands. The number of PM types mainly affects the pre-computation time spent on enumerating the number of feasible configurations that can be supported by each PM type. Since this enumeration is one-time effort and it is done in advance, the time spent on it is not counted towards the computation time for solving an instance of the VM placement problem. We will take advantage of the small number of VM types.}.

Let $\mathcal{T}^v$ denote the set of VM types. For each $u \in \mathcal{T}^v$, let $m_u$ be the total number of type-$u$ VMs that need to be placed. Let $\mathcal{T}^p$ denote the set of PM types. For each $v \in \mathcal{T}^p$, let $\mathcal{Y}_v$ be the set of all type-$v$ PMs. For different $v$, the sets $\mathcal{Y}_v$ are disjoint.

A {\em configuration} with an ID $t$ of a PM is a vector of non-negative integers, denoted by $\textbf{w}^t = (w_{u}^t)_{u \in \mathcal{T}^v}$, where each $w_{u}^t$ represents the number of type-$u$ VMs assigned to the PM in configuration $t$.
We say a configuration is {\em feasible} with respect to a PM if the configuration is supportable by the PM's resources, including allowing the disk anti-colocation constraints to be satisfied. For instance, suppose there are only $4$ VM types and suppose the vector $(3,5,4,0)'$ is a feasible configuration for a PM. That means the PM can support $3$ type-$1$ VMs, $5$ type-$2$ VMs, $4$ type-$3$ VMs and $0$ type-$4$ VMs simultaneously. For simplicity, we exclude the vector $0$ as a valid configuration, although this is not essential.

Since all PMs of the same type have the same amount of resources, a feasible configuration is also with respect to a PM type. Note that a configuration can be feasible to more than one PM types.

%

Suppose every configuration has a unique ID. For each PM type $v \in \mathcal{T}^p$, let $\mathcal{C}_v$ be the ID set of all the feasible configurations with respect to a type-$v$ PM. For this formulation, the configurations in $\mathcal{C}_v$ are assumed to be known (by preprocessing) and the number of them is assumed to be not too large, e.g., no more than hundreds of thousands. There are problem instances for which the assumptions hold (see Section \ref{sec:analysis} for the applicability of F2).
Note that the disk anti-collocation requirement must be satisfied in any feasible configuration. In the preprocessing step where we enumerate the feasible configurations for each PM type, we check the disk anti-collocation requirement.






For each PM type $v$, each PM $j \in \mathcal{Y}_v$ and each $t \in \mathcal{C}_v$, let $\gamma_{jt}$ be the $0$-$1$ assignment variable with $\gamma_{jt}=1$ if and only if PM $j$ is assigned to take the configuration $t$. The second formulation is as follows.
\begin{align}
\textbf{F2:}
\min_{\gamma,z} & \sum_{v \in \mathcal{T}^p, j \in \mathcal{Y}_v} \hat{c}_j z_j \label{eq:optobj_2} \\
\text{s.t.} \ \ & \sum_{t \in \mathcal{C}_v} \gamma_{jt} \leq 1, \ \ \ v \in \mathcal{T}^p, j \in \mathcal{Y}_v \label{eq:sumxij} \\
& \sum_{v \in \mathcal{T}^p} \sum_{j \in \mathcal{Y}_v} \sum_{t \in \mathcal{C}_v} \gamma_{jt} w^t_u \geq m_u, \ \ \ u \in \mathcal{T}^v \label{eq:allvmassigned} \\
& z_j \leq \sum_{t \in \mathcal{C}_v} \gamma_{jt}, \ \ \ v \in \mathcal{T}^p, j \in \mathcal{Y}_v \label{eq:z1_f2} \\
& B z_j \geq \sum_{t \in \mathcal{C}_v} \gamma_{jt}, \ \ \ v \in \mathcal{T}^p, j \in \mathcal{Y}_v \label{eq:z2_f2} \\
& \gamma_{jt}, z_j \in \{0,1\}, \ \ \ v \in \mathcal{T}^p, j \in \mathcal{Y}_v, t \in \mathcal{C}_v. \nonumber
\end{align}
The following explains some of the constraints:
\begin{itemize}
\item
(\ref{eq:sumxij}) ensures that, in a valid assignment, every PM must take at most one feasible configuration.
\item
(\ref{eq:allvmassigned}) guarantees that all the VMs are assigned to some PMs.
\item
(\ref{eq:z1_f2}) and (\ref{eq:z2_f2}) are the same as (\ref{eq:z1}) and (\ref{eq:z2}) in F1, which together ensure that $z_j = 1$ if and only if some $\gamma_{jt} = 1$.
\end{itemize}

Formulation F2 is visibly very different from formulation F1. It is useful if, for each PM type, the feasible configurations are enumerable and the number of them is not too large. More detailed analysis on the formulations is deferred till Section \ref{sec:analysis}.



\subsection{Formulation 3 -- Combined Formulation}


For some PM types, the number of feasible configurations may be too large for formulation F2 to be useful; i.e., F2 will have too many variables. For example, the l6 PM type in Table \ref{tab:PMtypes} has millions of feasible configurations (see Section \ref{sec:expsetup}). For other PM types, the number of feasible configurations may be small. For instance, if a PM does not have a lot of physical resources (e.g., it can support a total of $8$ vCPUs), then the number of feasible configurations is usually small. The s1 PM type in Table \ref{tab:PMtypes} has only 10 feasible configurations. We next consider a hybrid approach that combines formulations F1 and F2.


Let $\mathcal{P}_2$ be the set of PMs whose number of feasible configurations is not only enumerable, but also not too large (say, up to hundreds of thousands). Let $\mathcal{P}_1$ denote the set of the rest PMs, i.e., $\mathcal{P}_1 = \mathcal{P} \backslash \mathcal{P}_2$. The cutoff between the two sets should be based on computational experiences in the actual environment where our method is applied (see Section \ref{sec:analysis} for more discussion). The VM assignment to the PMs in $\mathcal{P}_2$ is done by choosing a configuration for each PM, as in formulation F2. The assignment to the PMs in $\mathcal{P}_1$ is done with the direct approach, i.e., by assigning VMs to PMs directly as in formulation F1. This combined approach is expected to work well if the number of PMs in $\mathcal{P}_1$ is not too large, say, up to several hundreds.


Let $\mathcal{T}^{p}_2$ be the set of PM types of all the PMs in the set $\mathcal{P}_2$.
For each $v \in \mathcal{T}^p_2$, let $\mathcal{Y}_v$ be the set of all type-$v$ PMs (which must be in $\mathcal{P}_2$), and let $\mathcal{C}_v$ be the ID set of all feasible configurations with respect to a type-$v$ PM. For a VM $i$, let $\tau(i)$ denote its type.
Let $\gamma_{jt}$ be the $0$-$1$ assignment variable with $\gamma_{jt}=1$ if and only if PM $j$ is assigned to take the configuration $t$.
The variables $x_{ij}$, $y_{ikjl}$ and $z_j$ are as in formulation F1. The combined formulation is as follows.

\begin{align}
\textbf{COMB:}
\min_{x,y,z,\gamma} & \sum_{j \in \mathcal{P}} \hat{c}_j z_j \label{eq:optobj_3} \\
\text{s.t.} \ \ & y_{ikjl} \leq x_{ij}, \ \ \ i \in \mathcal{V}, j \in \mathcal{P}_1, k \in R_i, l \in D_j \label{eq:mixyxp2} \\
& \sum_{j \in \mathcal{P}_1} \sum_{l \in D_j} y_{ikjl} = \sum_{j \in \mathcal{P}_1} x_{ij}, \ \ \ i \in \mathcal{V}, k \in R_i \label{eq:mustassign_mix} \\
& \sum_{j \in \mathcal{P}_1} x_{ij} \leq 1, \ \ \ i \in \mathcal{V} \label{eq:onePMonly_mix} \\
& \sum_{k \in R_i} y_{ikjl} \leq 1, \ \ \ i \in \mathcal{V}, j \in \mathcal{P}_1, l \in D_j \label{eq:diskexclusive_mix}
\end{align}

\begin{align}
& \sum_{i \in \mathcal{V}} \sum_{k \in R_i} \nu_{ik} y_{ikjl} \leq S_{jl}, \ \ \ j \in \mathcal{P}_1, l \in D_j  \label{eq:diskcap_mix} \\
& \sum_{i \in \mathcal{V}} \alpha_{i} x_{ij} \leq C_{j}, \ \ \ j \in \mathcal{P}_1 \label{eq:corecap_mix} \\
& \sum_{i \in \mathcal{V}} \beta_{i} x_{ij} \leq M_{j}, \ \ \ j \in \mathcal{P}_1  \label{eq:memcap_mix} \\
& z_j \leq \sum_{i \in \mathcal{V}} x_{ij}, \ \ \ j \in \mathcal{P}_1 \label{eq:z1_mix}\\
& B z_j \geq \sum_{i \in \mathcal{V}} x_{ij}, \ \ \ j \in \mathcal{P}_1 \label{eq:z2_mix} \\
& \sum_{t \in \mathcal{C}_v} \gamma_{jt} \leq 1, \ \ \ v \in \mathcal{T}^{p}_2, j \in \mathcal{Y}_v \label{eq:mixxpleq1} \\
& z_j \leq \sum_{t \in \mathcal{C}_v} \gamma_{jt}, \ \ \ v \in \mathcal{T}^{p}_2, j \in \mathcal{Y}_v
\label{eq:z3_mix}\\
& B z_j \geq \sum_{t \in \mathcal{C}_v} \gamma_{jt}, \ \ \ v \in \mathcal{T}^{p}_2, j \in \mathcal{Y}_v
\label{eq:z4_mix}\\
& \sum_{v \in \mathcal{T}^{p}_2} \sum_{j \in \mathcal{Y}_v} \sum_{t \in \mathcal{C}_v} \gamma_{jt} w^t_u \nonumber \\
& \quad + \sum_{i: \tau(i) = l} \sum_{j \in \mathcal{P}_1} x_{ij}
    \geq m_u, \ u \in \mathcal{T}^v
    \label{eq:VM_num_mix} \\
& x_{ij}, y_{ikjl} \in \{0,1\}, \ i \in \mathcal{V}, k \in R_i, j \in \mathcal{P}_1, l \in D_j \nonumber \\
& \gamma_{jt} \in \{0,1\}, \ \ \ v \in \mathcal{T}^{p}_2, j \in \mathcal{Y}_v, t \in \mathcal{C}_v \nonumber \\
& z_j \in \{0,1\}, \ \ \ j \in \mathcal{P}. \nonumber
\end{align}
The following explains some of the constraints:
\begin{itemize}
\item
(\ref{eq:mixyxp2})-(\ref{eq:z2_mix}) deal with direct VM assignment to the PMs in the set $\mathcal{P}_1$, which should be compared with (\ref{eq:diskconstonlyif})-(\ref{eq:z2}) in formulation F1.
More specifically,
(\ref{eq:mixyxp2}) ensures that the requested virtual disks for VM $i$ may be assigned to the
physical disks of PM $j$ only if VM $i$ is assigned to a PM $j$ in $\mathcal{P}_1$.
(\ref{eq:mustassign_mix}) ensures that every requested virtual disk must be assigned to exactly one physical disk only if VM $i$ is assigned to a PM $j$ in $\mathcal{P}_1$.
(\ref{eq:onePMonly_mix}) ensures that every VM is assigned to at most one PM in $P_1$.
(\ref{eq:diskexclusive_mix}) ensures that VM $i$ cannot have more than one of its virtual disks assigned to the same physical disk; (\ref{eq:mixyxp2}) and (\ref{eq:diskexclusive_mix}) together enforce the disk anti-colocation constraints.
(\ref{eq:diskcap_mix}) is the disk capacity constraint.
(\ref{eq:corecap_mix}) and (\ref{eq:memcap_mix}) are the resource capacity constraints posed by the number of vCPUs and the total memory size of each PM $j$.
(\ref{eq:z1_mix}) and (\ref{eq:z2_mix}) together ensure that $z_j=1$ if and only if $x_{ij}=1$ for some $i \in \mathcal{V}$, where $B$ is a large enough constant (it is enough to take $B = N$).

\item
(\ref{eq:mustassign_mix}) (\ref{eq:onePMonly_mix}) (\ref{eq:diskexclusive_mix})
are slightly different from their counterparts in formulation F1 -- (\ref{eq:mustassign}) (\ref{eq:onePMonly}) (\ref{eq:diskexclusive}) -- because each VM $i$ does not have to be assigned to a PM in the set $\mathcal{P}_1$.

\item
(\ref{eq:mixxpleq1})-(\ref{eq:z4_mix}) deal with VM assignment to the PMs in the set $\mathcal{P}_2$, which should be compared with formulation F2. (\ref{eq:mixxpleq1}) ensures that, in a valid assignment, every PM must take at most one feasible configuration. (\ref{eq:z3_mix}) and (\ref{eq:z4_mix}) together ensure that $z_j = 1$ if and only if some $\gamma_{jt} = 1$.

\item
The constraint (\ref{eq:VM_num_mix}) guarantees that all the VMs are assigned.
\end{itemize}

\subsection{Analysis of the Formulations}
\label{sec:analysis}


Formulations F2 and COMB are derived by reformulating the variables. They exploit special structures of the problem and define the variables very differently from what the obvious formulation, F1, does. As a result, the three formulations often have drastically different numbers of variables and constraints for the same problem instance. Our computational experiences have shown that the differences in computation time are often enormous. By counting the numbers of variables and constraints, it is often easy to see which formulation may be suitable and which are definitely impractical\footnote{The branch-and-bound algorithm used by the MIP solvers involves visiting the nodes on a branch-and-bound tree and solving a linear programming (LP) problem for each node visited. The numbers of variables and constraints are good predictors for the computation time of each LP problem. The number of constraints is a trickier criterion to use, as sophisticated MIP solvers often add more constraints in an attempt to ``tighten the constraints'' of the LP problems. The objective is to solve the original MIP problem faster by reducing the number of nodes visited on the branch-and-bound tree.}. For example, if the number of variables exceeds tens of millions, then the formulation clearly will be difficult to solve. Similarly, if the number of variables in formulation F2 exceeds that in formulation F1 by orders of magnitude, then F2 will most likely be more difficult to solve. If the number of constraints in formulation F1 exceeds that in formulation F2 by orders of magnitude, then F1 will most likely be more difficult to solve.

In order to use formulations F2 and COMB, the feasible configurations supported by each PM type need to be pre-computed, by enumeration. Since this enumeration is one-time effort and it is done in advance, the time spent on it is not counted towards the computation time for solving an instance of the VM placement problem. For each PM type, we only need to enumerate up to a million feasible configurations. If a PM type has more than a million feasible configurations, formulation F2 will not be solvable. We have to use formulation COMB and apply direct VM assignment to the PMs of that type.


\subsubsection{Formulation F1}

The total number of variables is dominated by the number of the $y_{ikjl}$ variables. That number is equal to the product of the total number of all the virtual disks in the problem with the total number of all the physical disks, i.e., $\bigl(\sum_{i \in \mathcal{V}} |R_i| \bigr) \times \bigl(\sum_{j \in \mathcal{P}} |D_j| \bigr)$. The number of constraints is also roughly the same.

Based on our computational experiences, when both numbers exceed hundreds of thousands, formulation F1 is impractical. When both numbers are below hundreds of thousands but above tens of thousands, F1 is likely solvable but may take a long time. When both numbers are less than tens of thousands, the formulation is often solvable fairly fast.

\subsubsection{Formulation F2}
\label{sec:analyzeF2}

The total number of variables is dominated by the number of $\gamma_{jt}$ variables, which is also the total number of configurations supported by all the PMs in the set $\mathcal{P}$, i.e., $\sum_{k \in \mathcal{T}^p} |\mathcal{Y}_k| |\mathcal{C}_k|$. If that number is greater than millions, the formulation will be either slow to solve or impossible to solve.
Otherwise, the formulation is generally faster to solve than formulation F1.
The number of constraints is roughly equal to $3$ times of the number of PMs, which is comparably small.

Formulation F2 is useful when the total number of configurations supported by all the PM types is not too large, e.g., under hundreds of thousands. It is generally easy to see when F2 is entirely impractical. For instance, a PM of a large type may have an exceedingly large number of feasible configurations, which will result in an exceedingly large number of variables and make formulation F2 impractical. An example is given in Section \ref{sec:expsetup}.


A small or moderate number of feasible configurations can happen if some combination of the following conditions is satisfied: (i) The number of VM types is small, e.g., dozens or less; (ii) a PM has a small capacity in at least one type of resources, e.g., 4-8 vCPUs; or (iii) there are policy-based restrictions ensuring that certain PM types are used only for a small number of specific VM types.

As an example of (ii), all the PM types with 8 vCPUs cannot accommodate any VM of the type i2.8xlarge, which demands 32 vCPUs (see Tables \ref{tab:VMtypes} and \ref{tab:PMtypes}). In general, for small or medium PM types, the number of feasible configurations is usually small because (1) a subset of the VM types are ruled out, and (2) for each remaining VM type, only a small number of such VMs can be assigned to a PM of the small or medium types due to resource scarcity.

As an example of (iii), a cloud provider may have a policy that the PMs of the large type are reserved for resource-intensive VM types. Such a policy is sensible for both economic and performance reasons, e.g., meeting the performance goals of high-value customers. More concretely, if each l2-type PM is only allowed to host the VM types with at least $8$ vCPUs requirement, then the number of feasible configurations reduces from more than $2 \times 10^{12}$ to $427$ (see Tables \ref{tab:VMtypes} and \ref{tab:PMtypes}).


\subsubsection{Formulation COMB}


For some large PM types, the number of feasible configurations may be large (say, more than hundreds of thousands). This is where formulation COMB is useful. We regard formulation COMB as one of the key contributions of the paper because it can treat large PMs separately by using direct VM assignment rather than using configurations. In the meantime, it treats the small or medium PM types by using configurations.

The total number of variables is roughly equal to the sum of the number of $y_{ikjl}$ variables and the number of $\gamma_{jt}$ variables in formulation COMB. The number of $y_{ikjl}$ variables is equal to $\bigl( \sum_{i \in \mathcal{V}} |R_i| \bigr) \times \bigl( \sum_{j \in \mathcal{P}_1} |D_j| \bigr)$, which should be compared with the case of F1. The number of $\gamma_{jt}$ variables is equal to $\sum_{k \in \mathcal{T}^p_2} |\mathcal{Y}_k| |\mathcal{C}_k|$, which is the number of feasible configurations supported by all the PMs in the set $\mathcal{P}_2$. The number of constraints is roughly equal to $\bigl( \sum_{i \in \mathcal{V}} |R_i| \bigr) \times \bigl( \sum_{j \in \mathcal{P}_1} |D_j| \bigr) + 3 |\mathcal{P}_2|$.

Thus, for formulation COMB to be effective, it is necessary that $\mathcal{P}_1$ contains a small number of PMs (e.g., no more than hundreds), and the PMs in $\mathcal{P}_2$ support a small to moderate number of feasible configurations, e.g., no more than hundreds of thousands. There is flexibility in setting the sets $\mathcal{P}_1$ and $\mathcal{P}_2$. Based on the above discussion, $\mathcal{P}_1$ should contain a small number of ``large" PMs, i.e., PMs with rich resources. With respect to the PMs in $\mathcal{P}_2$, a small or moderate number of feasible configurations can happen under the conditions (i)-(iii) given in Section \ref{sec:analyzeF2}. The above discussion provides a guideline for narrowing down the choices of $\mathcal{P}_1$ and $\mathcal{P}_2$. The final decision can be made based on computational experiences and by comparing the actual numbers of variables and constraints for different choices of $\mathcal{P}_1$ and $\mathcal{P}_2$, as the numbers can be easily computed.


Formulation COMB presents the most flexibility and applicability, because it contains formulations F1 and F2 as special cases. One can design good formulation COMB to speed up the computation or to solve larger instances.



\subsubsection{Summary of Formulation Analysis}

\begin{itemize}
\item 
In F1, the number of variables and the number of constraints are comparable. If F1 is impractical from the computation point of view, it is because both numbers are large.

\item 
F2 usually has a small number of constraints. If the number of variables is also small, which depends on the PM types in the problem instance, then F2 is likely to be faster to solve than F1. When F2 is impractical, it is usually because the number of variables is too large, which in turn is due to the presence of some resource-rich ("large") PM types.

\item 
If F2 is impractical, one can consider formulation COMB. The key is to decide the sets $\mathcal{P}_1$ and $\mathcal{P}_2$; the former contains the resource-rich PMs. In many problem instances, it is possible to drastically reduce the number of variables, as compared with F2, while only increase the number of constraints moderately. Then, COMB will be effective. When COMB is impractical, it is usually because there is a large number of resource-rich PMs, making the set $\mathcal{P}_1$ large. However, in practice, that is unlikely to happen often because cloud providers prefer to use commodity PMs for cost and ease-of-management reasons. Large PMs are rare, specialty items for special customers.

\item
There will be problem instances for which none of the formulations are practical. In those cases, one has to resort to other strategies, most likely using heuristic algorithms; but the solutions will not be optimal.

\end{itemize}


\section{Experiments}
\label{sec:experiments}

In this section, we will show problem instances and solve the three formulations using the MIP solver Gurobi \cite{Gurobi}. The main objective is to compare the computation time and show the vast differences among the three formulations. The results will reveal that formulation COMB can be used for large and complex problem instances.




\subsection{Setup}
\label{sec:expsetup}


We follow the VM and PM setup in Amazon's EC2 \cite{EC2Inst} as close as we can.
We take a subset of the allowed VM types (classes) of Amazon's EC2. Their resource requirements are shown in Table \ref{tab:VMtypes}. Cloud providers generally don't disclose the detailed capabilities of all their PMs. As discussed in Section \ref{sec:F2form}, the number of PM types is likely small. For the experiments, we assume the PM types are as shown in Table \ref{tab:PMtypes}. The amount of resources of each PM type is largely our guess based on the information revealed on Amazon's web site. The operation costs (in the 5th column) are also based on our estimate\footnote{The large cost increase when the number of disks exceeds 4 reflects the cost of running separate DAS (directed attached storage) devices.}. The costs are normalized, with the lowest operation cost chosen to be $100$. Since the problem is linear, it doesn't matter what the chosen normalization base cost is. If the base cost is chosen to be $\theta$ instead of $100$, the optimal cost is simply $\theta/100$ times of the optimal cost under the base cost $100$.




\begin{table}[t]
\caption{VM Types}
\label{tab:VMtypes}
\begin{center}
\begin{tabular}{|c|c|c|c|}
\hline
VM Type & vCPU & Memory (GiB) & Storage (all SSD; GB) \\
\hline
m3.medium &	1 &	3.75 &	1 $\times$ 4 \\
m3.large &	2 &	7.5 &	1 $\times$ 32 \\
m3.xlarge &	4 &	15 & 	2 $\times$ 40 \\
m3.2xlarge & 8 &	30 & 	2 $\times$ 80 \\
\hline
c3.large &	2 &	3.75 &	2 $\times$ 16 \\
c3.xlarge &	4 &	7.5 & 2 $\times$ 40 \\
c3.2xlarge &	8 &	15 &	2 $\times$ 80 \\
c3.4xlarge &	16 &	30 &	2 $\times$ 160 \\
c3.8xlarge 	& 32 &	60 & 	2 $\times$ 320 \\
\hline
r3.large &	2 &	15.25 &	1 $\times$ 32 \\
r3.xlarge &	4 &	30.5 &	1 $\times$ 80 \\
r3.2xlarge &	8 &	61 &	1 $\times$ 160 \\
r3.4xlarge &	16 &	122 &	1 $\times$ 320 \\
r3.8xlarge &	32 & 244 & 2 $\times$ 320 \\
\hline
i2.xlarge &	4 &	30.5 &	1 $\times$ 800 \\
i2.2xlarge &	8 &	61 &	2 $\times$ 800  \\
i2.4xlarge &	16 &	122 &	4 $\times$ 800  \\
i2.8xlarge &	32 &	244 &	8 $\times$ 800 \\
\hline
\end{tabular}
\end{center}
\end{table}

For Amazon EC2, each vCPU corresponds to a hyperthread of a physical core \cite{Field14}. In our experiments, we assume the PMs all support two hyperthreads per physical core. Hence, each physical core counts as 2 vCPUs. As an example, each Xeon E5-2680 processor has 8 cores and supports a total of 16 threads. A PM with one such processor offers 16 vCPUs.

For the PM types s1-s4 and m1-m5, we pre-computed all their feasible configurations. As stated earlier, the pre-computation step is a one-time effort and the required time is not counted toward the final computation time. In fact, for these PM types, the numbers of feasible configurations are quite small: s1--10, s2--36, s3--174, s4--174, m1--315, m2--2113, m3--4247, m4--4247, m5--3199. For the PM types l1-l6, we did not pre-compute their feasible configurations because they have much more resources and the numbers of feasible configurations are large. For example, the l6 PM type has millions of feasible configurations. Therefore, when PM types l1-l6 are involved in the experiments, we use formulation COMB instead of formulation F2. Experiment I and II are done with Gurobi-5.6.3 on a ThinkPad 220i laptop with 2 Intel i3 cores and 10G RAM.
The other experiments are done with Gurobi-6.5.2 on a ThinkPad 240 laptop with 2 Intel i7 cores and 8G RAM.
Gurobi is one of the highly regarded MIP solvers. Comparison results suggest that Gurobi is at least competitive against two other major commercial MIP solvers, CPLEX and XPRESS \cite{Mittelmann16}. All these commercial solvers are much faster than open-source alternatives.




\begin{table}[t]
\caption{PM Types}
\label{tab:PMtypes}
\begin{center}
\begin{tabular}{|c|c|c|c|c|}
\hline
PM Type & vCPU & Memory & Storage & Operation Costs \\
& & (GiB) & (all SSD; GB) & (normalized) \\
\hline
s1 &	8 &	16 & 1 $\times$ 256 & 100 \\
s2 &	8 &	32 &	1 $\times$ 512 & 120 \\
s3 &	8 &	64 & 	2 $\times$ 512 & 200 \\
s4 & 8 & 64 & 	4 $\times$ 512 & 300 \\
\hline
m1 &	16 &	32 &	2 $\times$ 512 & 600 \\
m2 &	16 &	64 & 4 $\times$ 512 & 700 \\
m3 &	16 &	128 &	4 $\times$ 1000 & 900 \\
m4 &	16 &	256 &	8 $\times$ 1000 & 1500 \\
m5 	& 16 &	256 & 	16 $\times$ 512 & 1800 \\
\hline
l1 &	32 &	256 &	4 $\times$ 1000 & 2500 \\
l2 &	48 &	512 &	8 $\times$ 1000 & 3500 \\
l3 &	64 &	1024 &	4 $\times$ 1000 & 5000 \\
l4 &	80 &	2048 &	16 $\times$ 1600 & 7000 \\
l5 &	120 & 4096 & 4 $\times$ 1000 & 9000 \\
l6 &	120 & 4096 & 24 $\times$ 1600 & 12000 \\
\hline
\end{tabular}
\end{center}
\end{table}

\subsection{Comparison with Greedy Randomized Heuristic Algorithm}

As a target for performance comparison, we developed our own heuristic algorithm. The heuristic algorithm is motivated by the general ideas of online heuristic algorithms \cite{Tang_ascalable, AAS14, Li20131222} but should achieve much lower costs than the latter due to two exhaustive search steps, which we will describe.

Imagine that VM requests arrive dynamically. An online randomized algorithm will assign a requested VM to some random PM one at a time in the arrival order of the VM requests. Note that, in our experiments, all the VMs to be placed are given together in a batch. Our greedy randomized algorithm first randomly permutes the list of all the requested VMs; this emulates the random arrival order of the VM requests. For each VM in the permuted list, an attempt is made to assign the VM to a PM. The greedy aspect is that, for assignment, the list of {\em used} PMs, which are those already with some assigned VMs, is checked first; if the VM cannot be assigned to any PM in the used list, then the list of unused PMs is checked. The greediness tends to lead to more VM consolidation. In scanning either PM list, the order of scanning is uniformly random to emulate random selection; the first PM in the list that can accommodate the VM is selected (first-fit)\footnote{For a large datacenter, scalable online algorithms cannot afford to search through all the used PMs or unused PMs for each VM request. A typical strategy is to randomly sample a few used PMs and, if that does not work out, pick randomly a unused PM with sufficient resources. Our heuristic algorithm should do better in the achievable objective value. A more sophisticated algorithm is to keep track of an ordered list of all the PMs according to certain criterion and assign the VM to the first one on the list that fits. In this case, exhaustive search is needed and scalability is limited. Our heuristic algorithm does not maintain an ordered list because there is no obvious criterion for the order due to the difficult disk anti-colocation requirement.}.

For each scanned PM, our heuristic algorithm checks whether it is possible to assign the currently considered VM to that PM. For vCPU or memory, all that is needed is to check whether the remaining number of vCPUs or the remaining memory is sufficient for the VM. For disk assignment, the algorithm exhaustively enumerates different disk assignment possibilities and uses the first one that is feasible\footnote{Checking the feasibility of disk assignment can be done by some standard assignment algorithm, which may be faster than enumeration but still takes some time. Either way, our heuristic algorithm has limited scalability, since in the worst case there is one disk assignment problem for every PM and for every VM request. But, it should achieve a lower cost than more scalable online randomized algorithms that do not check all the PMs for all possible disk assignment possibilities.}. If the disk assignment (for the currently considered VM and PM) cannot be done by the algorithm, it is because the assignment is infeasible.

\subsection{Main Results}

We summarize the computation time and achieved costs of all experiments in Table \ref{tab:summary_I_IV} and Table \ref{tab:summary_V}. For the randomized heuristic, each test case is repeated for $1000$ times and the average cost is reported. Note that, regardless of the VM-PM ratios ($N:M$) that we have experimented with, typically not all the PMs are used by the VMs in our solutions. The VMs are consolidated into fewer PMs because our optimization objective is to minimize the total operation cost of the active (i.e., used) PMs. Out of the $M$ PMs, only those used PMs will incur costs.

\subsubsection{Experiment I -- $70$ VMs and $50$ PMs}

We experimented with a problem of assigning $70$ VMs to $50$ PMs. The detailed setup is in Table \ref{tab:exp_setup_I_IV}. In this problem instance, only the small and medium types of PMs are used. Hence, we can compare formulations F1 and F2.
Judging by the VM and PM numbers, this is a small instance. However, formulation F1 involves $17950$ binary variables and $26120$ constraints, which make it non-trivial for any optimization software. Formulation F2 involves $51597$ variables and $168$ constraints. Formulations F1 and F2 are solved by Gurobi in $41.46$ and $0.31$ seconds, respectively, both yielding the optimal cost $4540$ with $24$ PMs used. The results demonstrate that if all the feasible configurations can be pre-computed and if their numbers are not too large, formulation F2 may be solved much faster than formulation F1. The reason is that F2 has much fewer constraints.
We also experimented with the randomized heuristic algorithm. The average cost obtained by the heuristic algorithm is $5431$, which is about $19.6\%$ higher than the optimal cost of $4540$.

\begin{figure}[t]
\begin{center}
\includegraphics[width=2.5in]{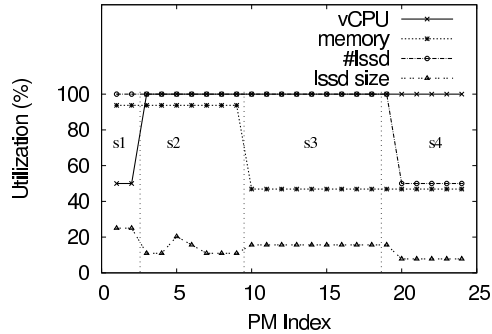}
\end{center}
\caption{Resource utilization resulted from formulation F1 for experiment I}
\label{fig:form1_set1}
\end{figure}
\begin{figure}[t]
\begin{center}
\includegraphics[width=2.5in]{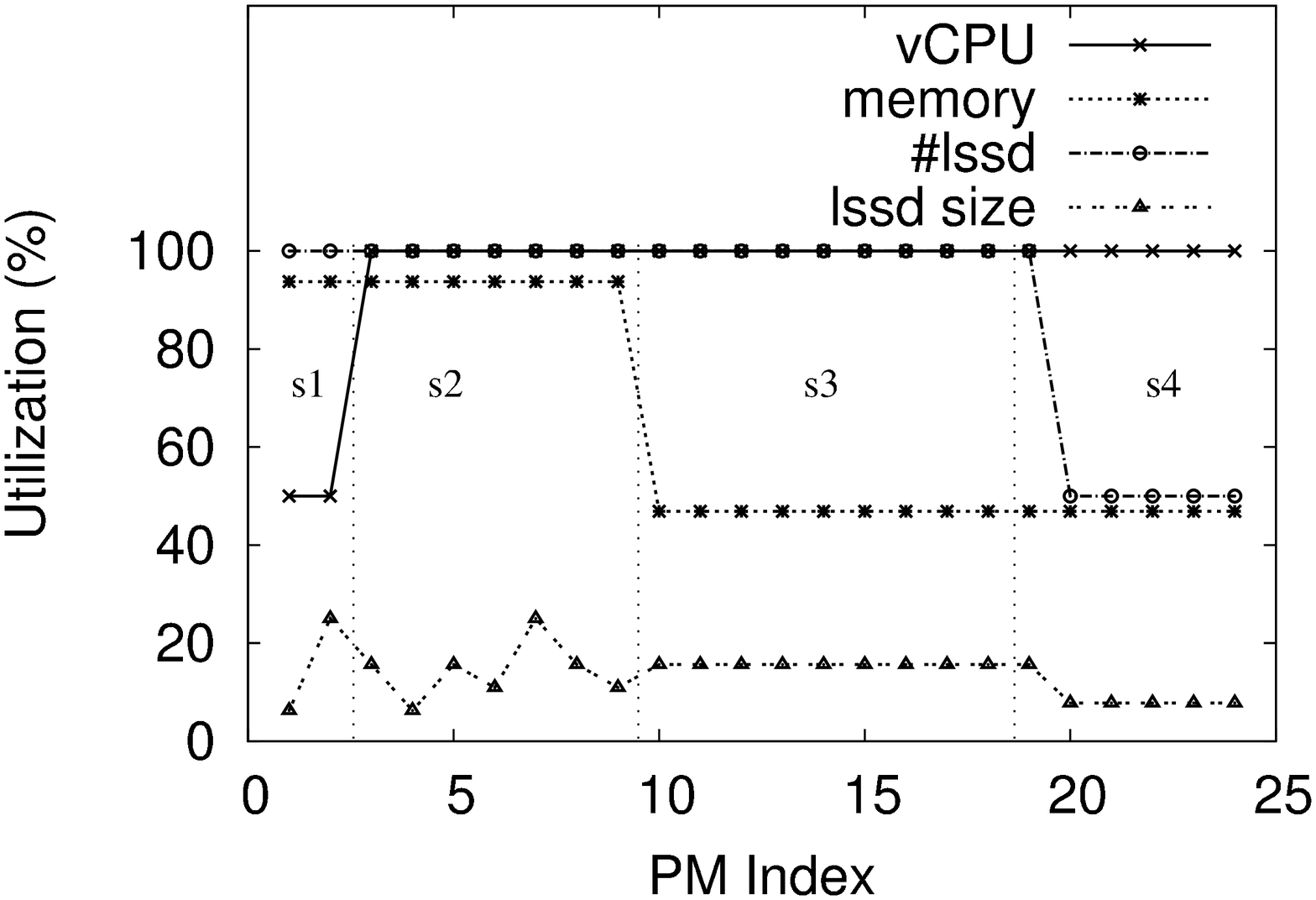}
\end{center}
\caption{Resource utilization resulted from formulation F2 for experiment I}
\label{fig:form2_set1}
\end{figure}

The obtained optimal solutions are useful for other purposes. For instance, they give indications on what resources are likely to be critical for different PM types. In Fig. \ref{fig:form1_set1} and Fig. \ref{fig:form2_set1}, we show the resource utilization of the PMs in the optimal solutions for formulations F1 and F2. For each resource and each PM, the utilization of that resource on the PM is defined as the ratio of the total requested amount by all the VMs assigned to that PM over the total available amount from that PM. For instance, suppose two VMs are assigned to a PM, and suppose each VM requires 4 vCPUs and the PM supports 8 vCPUs. Then, the vCPU utilization on the assigned PM is 100\%. Both solutions use $24$ PMs - s1: $2$; s2: $7$; s3: $10$; s4: $5$. Both solutions show very similar patterns of resource utilization. The vCPUs are critical resources for PM types s2, s3 and s4. The number of local disks (labeled as `\#lssd') is a critical resource for PM types s1, s2 and s3, in the sense that all those disks tend to have some virtual disks assigned to them. The memory utilization is high for PM types s1 and s2. The utilization of the physical disk capacity (labeled as `lssd size') is generally low (less than $30\%$). However, we should caution that these observations may change if the PMs have different resource configurations from what we are currently assuming.

\begin{figure}[t]
\begin{center}
\includegraphics[width=3in]{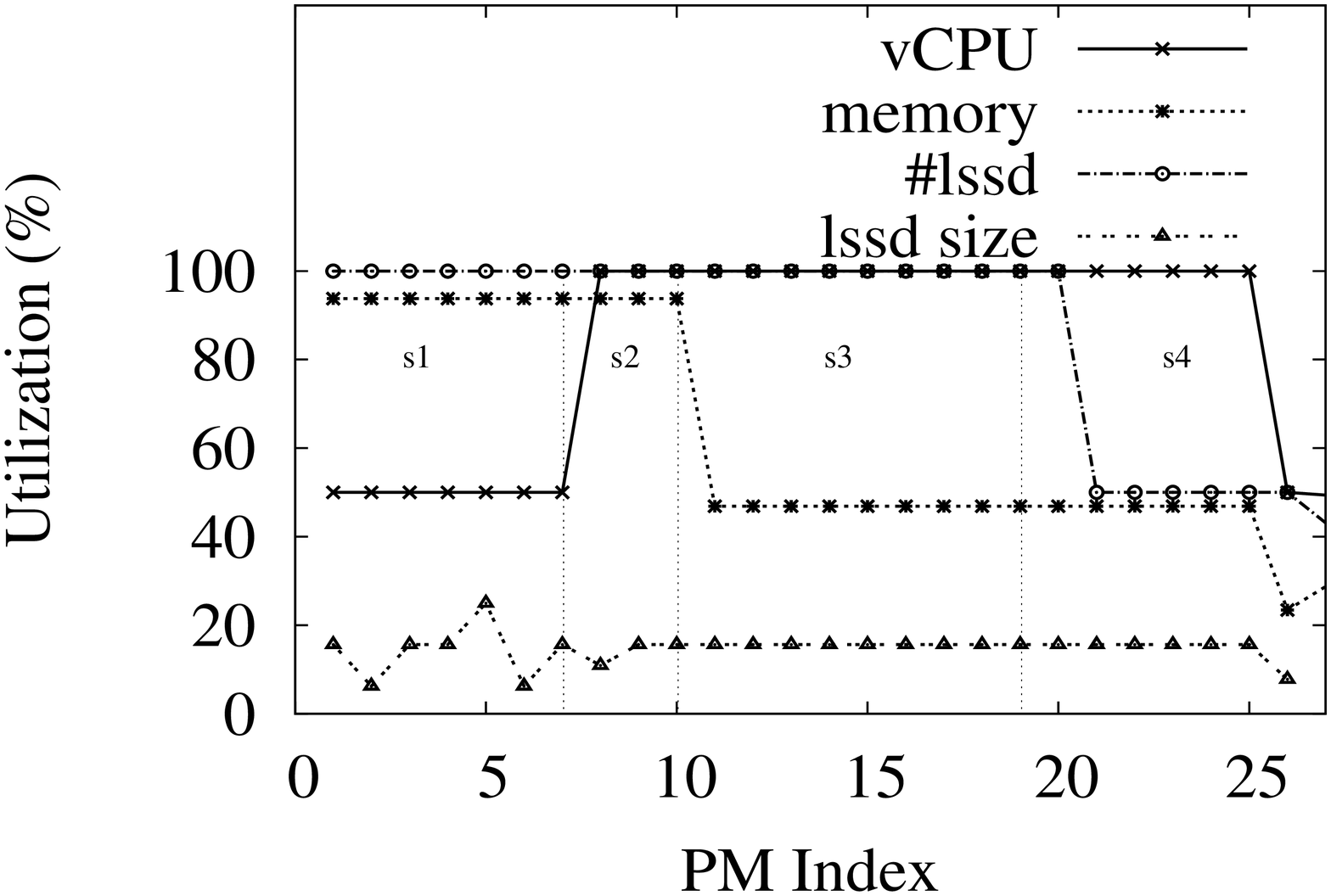}
\end{center}
\caption{Resource utilization resulted from randomized heuristic for experiment I}
\label{fig:random_set1}
\end{figure}

We also examined a solution produced by the randomized heuristic algorithm with a cost of $5160$ and $27$ active PMs. In Fig. \ref{fig:random_set1}, it shows that each PM type has a similar pattern of resource utilization as that of the optimal solutions obtained by F1 and F2. But the heuristic algorithm uses more s1-type PMs and fewer s2-type PMs. The s1 type has a larger vCPU-memory ratio compared with the s2 type. Therefore, the vCPU utilization of the s1 type is generally lower than the s2 type. The optimal solutions for F1 and F2 always use up all the available s2-type PMs. Meanwhile, the heuristic algorithm assigns more VMs to the s1-type PMs. Hence, the overall performance of the heuristic algorithm is worse.

With the objective of minimizing the total operation cost of the PMs, the optimal solution always seeks to improve the resource utilization of the active PMs. Hence, for almost every active PM, at least one resource is fully utilized. If that is not the case for some active PM, it is because there are no more unassigned VMs that can fit in that PM.

\subsubsection{Experiment II -- $77$ VMs and $70$ PMs}

In this experiment, $77$ VMs will be assigned to $70$ PMs. Although the numbers of VMs and PMs are not so different from the previous problem instance, the mixes of the VM and PM types are quite different (see Table \ref{tab:exp_setup_I_IV}). Here, we have a fuller mix of almost all types of VMs and PMs. Formulation F2 is impractical for this instance, because the numbers of configurations for some large PM types are too great to be pre-computed within a reasonable amount of time. Hence, we compare formulation COMB with formulation F1.
Formulation F1 has $55380$ binary variables and $80825$ constraints, quite a bit larger than the previous problem instance. Formulation COMB has
$97610$ binary variables and $37538$ constraints. Formulation F1 takes Gurobi about $2778$ seconds (about $46$ minutes) to solve, which is much longer than for the previous instance. Formulation COMB takes $955$ seconds, which is about one third of the time for Formulation F1. We see that, for this problem instance, formulation COMB has a modest computation time advantage over F1, but a great advantage over F2.
The optimal assignment has a cost of $45300$ and the average cost reported by the heuristic is $51102$.

With respect to resource utilization, the vCPUs and the number of disks are still critical resources for most PM types. The memory utilization is very high for more than half of the PMs. The disk capacity is often less than $50\%$ utilized for all PM types other than m3, l1 and l5.



\subsubsection{Experiment III \& IV -- around $1000$ VMs and $1000$ PMs}
We further experimented with a much larger example, where $1000$ VMs are to be assigned to $1000$ PMs of different types. The mixes of VMs and PMs are described in the part about Experiment III in Table \ref{tab:exp_setup_I_IV}. The results are summarized in Table \ref{tab:summary_I_IV}.

For this experiment, formulation F1 fails to finish running due to the large number of variables and constraints. Formulation F2 has $1099900$ binary variables and $3018$ constraints. It took $6.84$ seconds to solve. We find that the vCPUs and the number of disks are still critical resources for most PM types.

In the setup of Experiment III, there are not any large VMs and PMs. For Experiment IV, we added 10 large VMs and 12 large PMs, as shown in Table \ref{tab:exp_setup_I_IV}. The experiment emulates a scenario where an enterprise customer deploys around $1000$ VMs for its workforce. Most of the VMs are ordinary (not very powerful), and they are intended to be used by regular office workers. But, some large VMs are needed by power users or larger servers, and they must be put on large PMs. For this experiment, formulation F2 is impractical and formulation F1 fails to finish running. Formulation COMB has $238932$ binary variables and $124695$ constraints. It takes $2336$ seconds to solve.

For the setup of Experiment III, we ran $10$ additional experiments on random combinations of the VM and PM types. More specifically, we kept $1000$ VMs and $1000$ PMs. We randomly assigned the VMs into the m3 types and c3 types, and randomly assigned the PMs into the s types and m types. The results of using formulation F2 are reported in Table \ref{tab:summary_random_combination}. There are $3018$ constraints for each combination, and around $2,000,000$ binary variables, depending on the specific PM types. The computation time is around dozens of seconds.

\subsubsection{Experiment V, VI and VII -- Policy-Based Examples}

As discussed earlier, the number of feasible configurations for the large PM types can be very large, which limits the usefulness of formulations F2 and COMB. In reality, for economic or performance reasons, datacenters may have policies that large PMs are reserved for VMs that require a large amount of resources (see Section \ref{sec:analysis} for more discussion). Under such policies, the number of feasible configurations for large PMs can be drastically reduced, making formulation F2 and COMB more widely applicable. This group of experiments demonstrate the above points. The parameters for Experiment V, VI and VII are given in Table \ref{tab:exp_setup_V}. Experiment V has 6020 VMs and 2012 PMs, a fairly large deployment. Experiment VII has 7500 VMs and 6000 PMs, an even larger deployment; moreover, all types of VMs and PMs are involved. Experiment VI is a smaller deployment, but has some large VM and PM types.



Suppose a datacenter has the following policy restrictions for large PMs (l-type), and suppose there are no restrictions for the small (s-type) or medium (m-type) PM types.
\beit
\item l1 is restricted to: m3.xlarge, m3.2xlarge, c3.xlarge, c3.2xlarge, c3.4xlarge,
c3.8xlarge, r3.xlarge, r3.2xlarge, r3.4xlarge, r3.8xlarge, i2.xlarge, i2.2xlarge, i2.4xlarge, i2.8xlarge.
\item l2 and l3 are restricted to: m3.2xlarge, c3.2xlarge, c3.4xlarge,
c3.8xlarge, r3.2xlarge, r3.4xlarge, r3.8xlarge, i2.2xlarge, i2.4xlarge, i2.8xlarge.
\item l4, l5 and l6 are restricted to: c3.4xlarge,
c3.8xlarge, r3.4xlarge, r3.8xlarge, i2.4xlarge, i2.8xlarge.
\eeit

With the above policy, the number of configurations for each l-type PM is small and we can use formulation F2 to solve the problems in Experiment V, VI and VII. The results are summarized in Table \ref{tab:exp_setup_V}. All the problems can be solved very quickly under formulation F2, from under 1 second to 261 seconds. In contrast, without the policy-based restrictions, the two problems in Experiment V and VII cannot be solved using any of the three formulations. The problem in Experiment VI cannot be solved using formulation F1 or formulation F2.




\subsubsection{Experiments with Different VM-PM Ratios}
\begin{figure}[t]
\begin{center}
\includegraphics[width=3.5in]{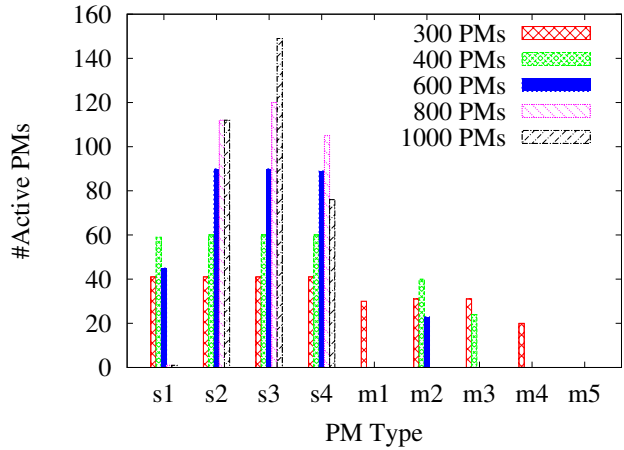}
\end{center}
\caption{Number of active PMs of each PM type}
\label{fig:skewed_PM}
\end{figure}

\begin{figure}[t]
\begin{center}
\includegraphics[width=3.5in]{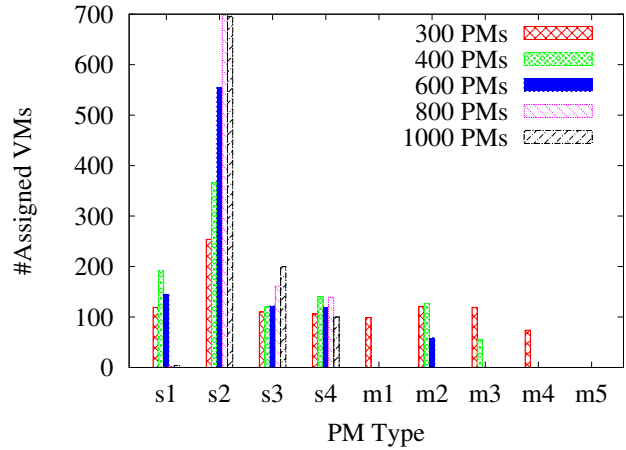}
\end{center}
\caption{Number of assigned VMs of each PM type}
\label{fig:skewed_VM}
\end{figure}

In the cloud environment, the ratio between the VMs and PMs may vary. In this set of experiments, we start with the setup of Experiment III but vary the number of PMs. More specifically, there are $300$, $400$, $600$, $800$ and $1000$ PMs in each experiment, where the proportion of each PM type is the same as that of Experiment III. We keep the VM setup of Experiment III unchanged, which has $1000$ VMs. The experiment with $200$ PMs is not feasible; hence we do not report its result.
We summarize the performance results in Table \ref{tab:skewed_ratio}, and plot the number of active PMs and the number of assigned VMs of each PM type in Fig. \ref{fig:skewed_PM} and Fig. \ref{fig:skewed_VM}, respectively. The results show that for the m3-type VMs, the PM types s2, s3, and s4 are more cost-efficient compared with the s1-type and the m-type PMs. Thus, with more available PMs from each PM type, the optimal solution shifts the VM assignments to the PM types s2, s3 and s4. Though more PMs need to be turned on, the overall cost is decreased from $127,120$ to $66,040$.

\begin{table}[t]
\caption{VM and PM Setup for Experiments I, II, III and IV}
\label{tab:exp_setup_I_IV}
\begin{center}
\begin{tabular}{|c|c|c|c|c|c|c|c|c|c|}
\hline
VM Type & \multicolumn{4}{|c|}{No. of VMs} & PM Type & \multicolumn{4}{|c|}{No. of PMs} \\
\hline
Experiment & I & II & III & IV  & Experiment & I & II & III & IV  \\
\hline
m3.medium &	36 & 5 & 500 & 500  & s1 & 7 & 5 & 150 & 150  \\
m3.large &	14 & 5 & 200 & 200  & s2 & 7 & 5 & 150 & 150 \\
m3.xlarge &	10 & 5 & 150 & 150 & s3 & 10 & 5 & 150 & 150 \\
m3.2xlarge & 10 & 5 & 150 & 150  & s4 & 7 & 5 & 150 & 150 \\
\hline
c3.large & & 5 &  & 2  & m1 & 5 & 5 & 100 & 100  \\
c3.xlarge & & 5 &  & 2  & m2 & 5 & 5 & 100 & 100  \\
c3.2xlarge & & 5 &  & 2  & m3 & 5 & 5 & 100 & 100 \\
c3.4xlarge & & 5 &  & 2 &  m4 & 2 & 5 & 50 & 50 \\
c3.8xlarge 	& & 5 &  & 2 & m5 & 2 & 5 & 50 & 50 \\
\hline
r3.large & & 5 &  &  &  l1 &  & 5 & & 2 \\
r3.xlarge &	& 5 &  &   & l2 &  &	5  & & 2 \\
r3.2xlarge & & 5 &  &   & l3 &  & 5 & & 2 \\
r3.4xlarge & & 5 &  &   & l4 &  & 5 & & 2 \\
r3.8xlarge & & 5 &  &  &  l5 &   & 5 & & 2 \\
       & & & & & l6 &  &  & & 2 \\
\hline
i2.xlarge & & 2  & & &	 & & & & \\
i2.2xlarge & & 2 & & &  & & & & \\
i2.4xlarge & & 3 & & &  & & & & \\
\hline
\end{tabular}
\end{center}
\end{table}

\begin{table}[t]
\caption{VM and PM Setup for Experiment V, VI and VII}
\label{tab:exp_setup_V}
\begin{center}
\begin{tabular}{|c|c|c|c|c|c|c|c|}
\hline
VM Type & \multicolumn{3}{|c|}{No. of VMs} & PM Type & \multicolumn{3}{|c|}{No. of PMs} \\
\hline
Experiment & V & VI & VII & Experiment & V & VI & VII \\
\hline
m3.medium &	0 & 0 & 1875 & s1 & 300 & 0 & 900\\
m3.large &	4000 & 0 & 750 & s2 & 300 & 0 &	900\\
m3.xlarge &	2000 & 0 & 563 & s3 & 300 & 0 &	900\\
m3.2xlarge & 0 & 15 & 562 & s4 & 300 & 0 & 900\\
\hline
c3.large &	0 &	0 & 600 & m1 & 200 & 10& 450\\
c3.xlarge &	0 &	0 & 600 & m2 & 200 & 10& 375\\
c3.2xlarge & 0 & 0 & 150 & m3 &	200 & 10& 375\\
c3.4xlarge & 3 & 15 & 75 & m4 & 100 & 10& 375\\
c3.8xlarge 	& 3 & 15 & 75 & m5 & 100 & 10& 375\\
\hline
r3.large &	0 & 0 &	600 & l1 &	2 &	5& 75\\
r3.xlarge &	0 &	0 & 600 & l2 &	2 &	5& 75\\
r3.2xlarge & 0 & 0 & 150 & l3 &	2 &	5& 75\\
r3.4xlarge & 3 & 0 & 150 & l4 &	2 &	5& 75\\
r3.8xlarge & 3 & 15 & 75 & l5 & 2 & 5& 75\\
& & & & l6 & 2 & 2 & 75 \\
\hline
i2.xlarge &	0 & 0 &	300 & & & &	\\
i2.2xlarge & 3 & 15 & 300 & & & &	\\
i2.4xlarge & 3 & 15 & 75 & & & &	\\
i2.8xlarge & 2 & 15 & 75 & & & &	\\
\hline
Total & 6020 & 105 & 7500 & Total & 2012 & 80 & 6000 \\
\hline
\end{tabular}
\end{center}
\end{table}

\begin{table}[t]
\caption{Summary of Computation Time (seconds) and Achieved Costs: Experiment I, II, III and IV}
\label{tab:summary_I_IV}
\begin{center}
\begin{tabular}{|c|c|c|c|c|}
\hline
Experiment & I & II & III & IV\\
\hline
Run Time of F1 & 41.46 & 2278 & N/A & N/A \\
\hline
Run Time of F2 & 0.31 & N/A & 6.84 & N/A \\
\hline
Run Time of COMB & N/A & 955 & N/A & 2336 \\
\hline
Cost by Optimization & 4540 & 45300 & 66040 & 73340 \\
\hline
Cost by Heuristics & 5431 & 51102 & 78628 & 85930\\
\hline
\end{tabular}
\end{center}
\end{table}

\begin{table}[t]
\caption{Summary of Computation Time (Seconds): Random Combination of VM Types and PM Types, $1000$ VMs and $1000$ PMs.}
\label{tab:summary_random_combination}
\begin{center}
\begin{tabular}{|c|c|c|c|}
\hline
Combination & Run Time (seconds) & \#Constraints & \#Binaries\\
\hline
1 & 6.84 & 3018 & 1099900 \\
\hline
2 & 20.47 & 3018 & 1512195 \\
\hline
3 & 13.71 & 3018 & 1804207 \\
\hline
4 & 9.45 & 3018 & 1865270 \\
\hline
5 & 14.99 & 3018 & 1719534 \\
\hline
6 & 13.18 & 3018 & 2244564 \\
\hline
7 & 11.48 & 3018 & 2090610 \\
\hline
8 & 17.59 & 3018 & 2506636 \\
\hline
9 & 16.35 & 3018 & 2009020 \\
\hline
10 & 26.50 & 3018 & 2260166 \\
\hline
\end{tabular}
\end{center}
\end{table}

\begin{table}[t]
\caption{Summary of Results: Experiment V, VI and VII}
\label{tab:summary_V}
\begin{center}
\begin{tabular}{|c|c|c|c|}
\hline
Experiment & V & VI & VII \\
\hline
Run Time of F2 & 15s & 0s & 261s \\
\hline
Cost by Optimization & 657200 & 170000 & 1046271\\
\hline Cost by Heuristics & 666805 & 184710 & 1851922 \\
\hline
\#Varialbes of F2 & 2207686 & 56 & 3747950 \\
\hline
\#Constraints of F2 & 6054 & 633 & 4018 \\
\hline
\end{tabular}
\end{center}
\end{table}

\begin{table}[t]
\caption{Summary of Tests with Skewed VM-PM Ratios; \#VM = $1000$}
\label{tab:skewed_ratio}
\begin{center}
\begin{tabular}{|c|c|c|c|c|c|}
\hline
\#PM & $300$ & $400$ & $600$  & $800$ & $1000$\\
\hline
Cost by Optimization & 127120 & 92700 & 76100 & 69040 & 66040 \\
\hline
Cost by Heuristics & 128370 & 106091 & 101333 & 86380 & 78628 \\
\hline
Active PMs & 276 & 303 & 337 & 338 & 338 \\
\hline
\end{tabular}
\end{center}
\end{table}


\section{Conclusions and Discussions}
\label{sec:conclusion}

In this paper, we examine the approach of using MIP formulations and algorithms for a special VM placement problem, which has difficult disk anti-colocation constraints. One of the key challenges is the potentially long computation time of MIP algorithms. We explore how different problem formulations -- by redefining variables -- can help to reduce the computation time. Our main effort is on developing the non-obvious formulations F2 and COMB. For a given problem instance, the three formulations often have drastically different numbers of variables and constraints, and the differences in computation time are often enormous. For many problem instances, it is easy to see which formulation may be suitable and which are definitely impractical, by counting the number of variables and the number of constraints. In the end, all three formulations can be useful, but for different problem instances. They all should be kept in the toolbox for tackling the problem. Out of the three, formulation COMB is especially flexible and versatile, and it can solve large problem instances.



The approach used by the paper is extensible to other datacenter resource management problems. For a given problem, different formulations likely exist and they can have very different computation time; which formulation has the least computation time depends on the problem instances. Thus, it is important to explore different formulations and select suitable ones for different instances.

Even with proper formulations, MIP algorithms can only solve what might be considered small to medium problem instances in our application setting, good enough for perhaps 1,000 -- 10,000 PMs. To model problems for a large datacenter in its entirety, an MIP formulation may involve trillions of variables and/or constraints, and there is no hope to solve them optimally within acceptable time. In such cases, we show in \cite{XTFC17} that a hierarchical decomposition heuristic can be effective. The decomposition method breaks a large, hard problem into many independent subproblems, which can be solved in parallel by separate control servers. Each of the subproblems can be made sufficiently small and solvable quickly using MIP algorithms. The material of this paper is relevant to those subproblems.

Finally, the VM placement problems encountered in practice will likely contain multiple difficult components, expressed by different sets of constraints or requirements. Each of the difficult components may require different techniques to cope with. A complete solution will need to combine those techniques together. This paper examines one such difficult component and provides one class of techniques, which can be used as a building block for solving problems encountered in practice.

\section{Acknowledgments}

X. Zheng was supported by the Shanghai Committee of Science
and Technology, China (Grant No. 14510722300),
and the Youth Innovation Promotion Association, CAS.

\bibliographystyle{elsarticle-num}
\bibliography{cloud}

\end{document}